\documentclass[sigconf, screen]{acmart}

\copyrightyear{2026}
\acmYear{2026}
\setcopyright{cc}
\setcctype{by}
\acmConference[SCF '26]{Proceedings of the 11th ACM Symposium on Computational Fabrication}{December 06--08, 2026}{Tokyo, Japan}
\acmBooktitle{Proceedings of the 11th ACM Symposium on Computational Fabrication (SCF '26), December 06--08, 2026, Tokyo, Japan}
\acmDOI{10.1145/3828611.3845638}
\acmISBN{979-8-4007-2821-1/2026/12}

\begin{document}
\graphicspath{{figures/}}
\title{Computational Design of Free-Form Lampshades}

\author{Hiro Nakasuji}
\email{sujihironaka@g.ecc.u-tokyo.ac.jp}
\orcid{0009-0008-8651-0285}
\affiliation{
  \institution{The University of Tokyo}
  \city{Tokyo}
  \country{Japan}
}
\author{Soma Yokota}
\email{yokota@cs.tsukuba.ac.jp}
\orcid{0009-0005-2054-3111}
\affiliation{
  \institution{University of Tsukuba}
  \city{Ibaraki}
  \country{Japan}
}
\author{Yutaka Ohtake}
\email{ohtake@race.t.u-tokyo.ac.jp}
\orcid{0000-0002-1368-9172}
\affiliation{
  \institution{The University of Tokyo}
  \city{Tokyo}
  \country{Japan}
}
\author{Yuki Koyama}
\email{koyama@pe.t.u-tokyo.ac.jp}
\orcid{0000-0002-3978-1444}
\affiliation{
  \institution{The University of Tokyo}
  \city{Tokyo}
  \country{Japan}
}

\begin{abstract}
Conventional lampshades are predominantly restricted to simple geometries, owing to the difficulty of controlling optical functions across complex shapes.
We present a computational design method that enables 3D-printable lampshades with complex free-form shapes to achieve controlled illumination, such as uniform luminance.
Given an arbitrary shape, our method determines the internal cavity geometry and the number and placement of lights in two stages: an initialization stage fixes the number of lights and establishes their initial positions and the initial cavity geometry, after which gradient-based optimization jointly refines the cavity geometry and light positions via differentiable rendering subject to geometric constraints.
We calibrate the material parameters of the translucent shade material prior to optimization to improve the agreement between simulation and the fabricated result.
We validate our approach by 3D-printing five lampshade designs of varying geometric complexity and demonstrate that the fabricated lampshades achieve reasonably uniform luminance.

\end{abstract}


\begin{teaserfigure}
  \includegraphics[width=\textwidth]{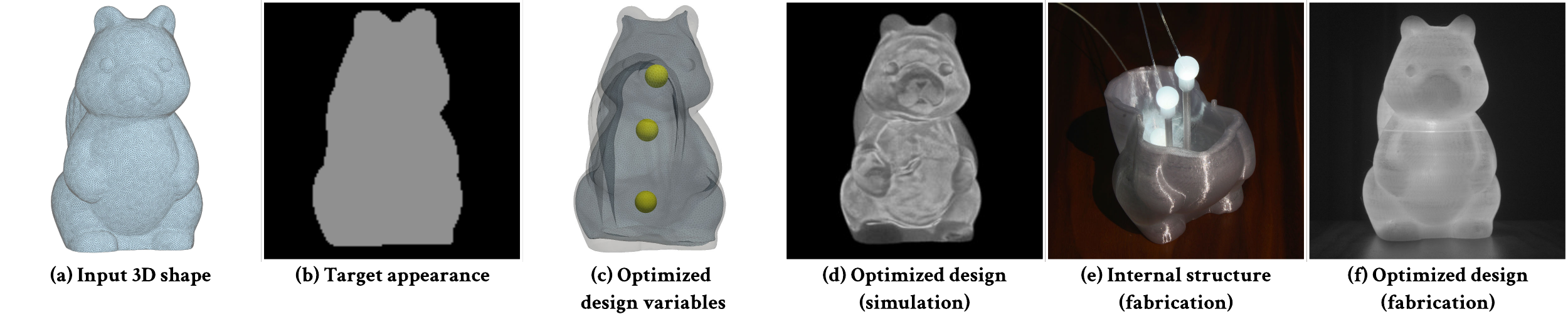}
  \caption{Design pipeline. (a) Input 3D shape for which a lampshade is to be designed. (b) Target appearance image specifying the desired surface appearance: here, uniform luminance across the silhouette. (c) Optimized internal cavity geometry and light placement. (d) Simulation rendering of the optimized design. (e) Internal structure of the fabricated lampshade, revealing the optimized internal cavity and fiber-fed sphere lights. (f) The fabricated result achieves the desired appearance.}
  \Description{Six panels showing the design pipeline: (a) an input 3D squirrel mesh, (b) a uniform gray silhouette representing the target appearance, (c) the optimized internal cavity geometry and light placement shown inside the transparent squirrel mesh with sphere lights visible, (d) a simulation rendering of the optimized design, (e) a photograph of the interior of the fabricated lampshade half revealing the internal cavity and fiber-fed sphere lights mounted on printed struts, and (f) a photograph of the completed fabricated lampshade.}
  \label{fig:teaser}
\end{teaserfigure}

\maketitle

\section{Introduction}

\begin{figure}[t]
  \centering
  \includegraphics[width=\linewidth]{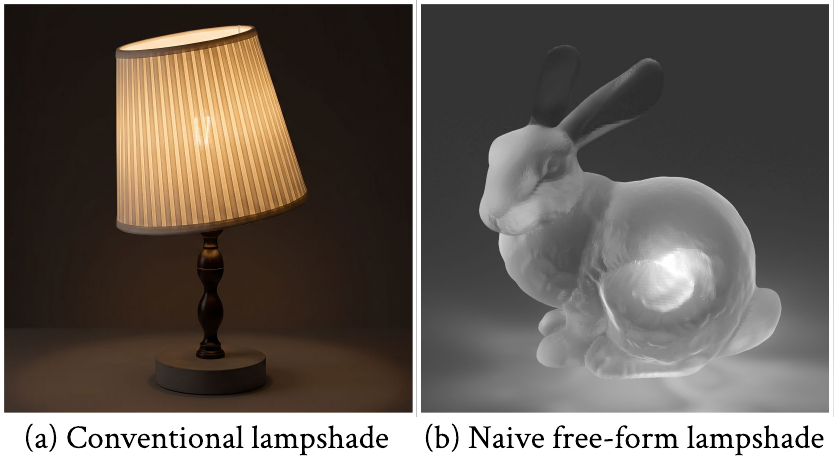}
  \caption{Motivation. (a) Conventional lampshades are restricted to simple geometries such as cylinders. (b) A free-form lampshade with a complex geometry is aesthetically appealing but, without a principled design method, fails to achieve the desired optical function (e.g., uniform luminance), producing a bright hot spot and dark areas elsewhere.}
  \Description{Two panels: (a) a photograph of a conventional cylindrical lampshade with a fabric shade on a decorative stand, and (b) a rendered bunny-shaped lampshade emitting light with uneven surface luminance concentrated toward the center.}
  \label{fig:motivation}
\end{figure}

Lampshades define the character of an interior space by diffusing and reflecting light to create the desired atmosphere.
Beyond their functional role, they carry cultural significance as handcrafted objects and works of decorative art, and there is growing demand in the lighting industry for designs that harmonize expressive form with precise optical behavior.
However, conventional lampshades are predominantly restricted to simple geometries such as cylinders and cones (\autoref{fig:motivation}(a)).
This limitation is fundamentally tied to the difficulty of predicting and controlling illumination effects for complex shade geometries (\autoref{fig:motivation}(b)).

We address the challenge of designing lampshades with complex free-form shapes using a computational approach. Recent advances in differentiable rendering have made it possible to compute gradients of image-based objectives with respect to geometric and material parameters, enabling automatic inverse design of complex optical systems via gradient descent~\citep{nimier2019mitsuba,jakob2022dr}.
This capability has been applied to a range of computational fabrication problems. However, designing functional free-form lampshades poses distinct technical challenges. The shade material exhibits translucent volumetric light transport, and the visible shape is fixed and cannot be altered to achieve the desired illumination.

We present a computational design method that enables 3D-printable lampshades with complex free-form shapes to achieve controlled illumination (\autoref{fig:teaser}). As lampshades are typically desired to emit light uniformly, we focus on uniform luminance as the design target throughout this work.
Our method takes an arbitrary shape as a fixed input and optimizes two quantities hidden within it: the geometry of the internal cavity and the number and placement of lights.
An initialization stage first determines the number of lights and establishes their initial positions together with the initial cavity geometry. Gradient-based optimization via differentiable rendering then jointly refines the cavity geometry and light positions subject to geometric constraints.
In addition, we calibrate the material parameters of the translucent shade material prior to optimization to improve the agreement between simulation and physical appearance.
We validate our approach across five shapes of varying geometric complexity through 3D printing, demonstrating that our method substantially improves luminance uniformity over a naive design and achieves reasonably uniform luminance even on geometrically complex lampshades.
The contribution of this work is a complete design-to-fabrication pipeline for translucent free-form lampshades, integrating differentiable rendering for joint optimization of internal geometry and light placement, physical material calibration to bridge the simulation-to-fabrication gap, and fabrication validation on physical lampshades.
\section{Related Work}

\begin{figure*}[t]
  \centering
  \includegraphics[width=\textwidth]{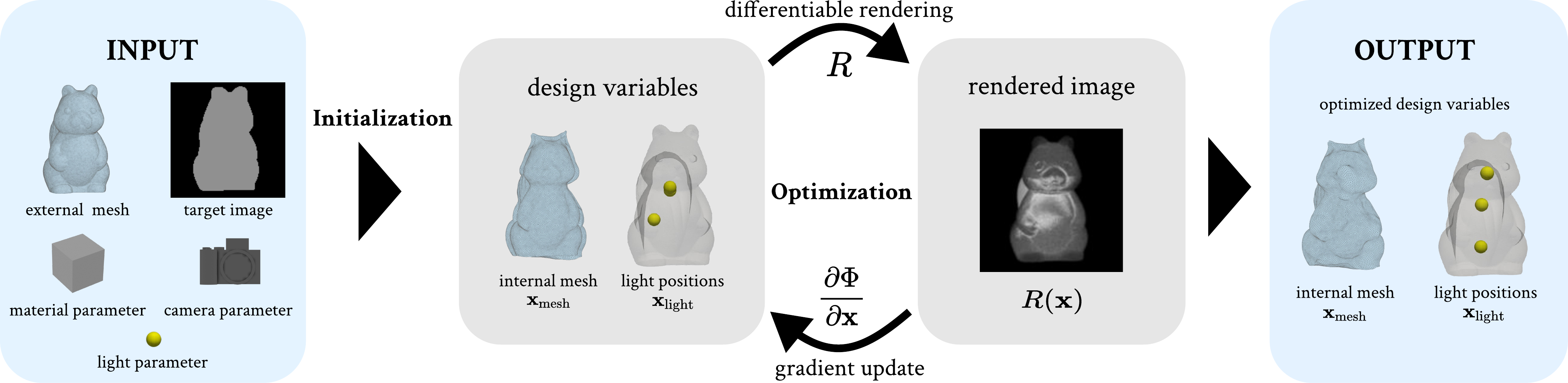}
  \caption{Overview of the design pipeline. Given an external mesh, a target luminance image, and material, camera, and light parameters, the initialization stage fixes the initial design variables: the vertex positions of the internal cavity mesh $\mathbf{x}_\mathrm{mesh}$ and the light positions $\mathbf{x}_\mathrm{light}$, whose count and initial placement are jointly determined via initialization. The optimization stage then minimizes the objective function $\Phi$ by iterating differentiable rendering $R$ and gradient updates $\partial\Phi/\partial\mathbf{x}$, yielding the optimized internal geometry and light placement as output.}
  \Description{A diagram of the design pipeline. On the left, the inputs are listed: an external mesh, a target luminance image, and material, camera, and light parameters. An arrow labeled Initialization leads to the design variables: an internal cavity mesh x\_mesh and a light configuration x\_light, visualized as a 3D squirrel model with lights inside the cavity. A loop labeled Optimization connects the design variables to a rendered image R(x) via differentiable rendering R (top arrow), and back to the design variables via gradient update partial-Phi over partial-x (bottom arrow). On the right, the output shows the optimized design variables with the redistributed lights and deformed internal cavity.}
  \label{fig:overview}
\end{figure*}

\subsection{Functional Fabrication with Lights}
Computational fabrication has been applied to control how light interacts with physical objects, with prior work focusing on specific optical phenomena such as refraction, reflection, or occlusion.

One line of research achieves desired optical effects by optimizing the geometry of the fabricated object itself.
\citet{finckh2010geometry}, \citet{papas2011goal}, and \citet{schwartzburg2014high} optimize the geometry of transparent objects to produce target caustic patterns via refraction and reflection.
\citet{mitra2009shadow} instead exploit occlusion, optimizing 3D voxel geometries that cast distinct silhouette shadows from multiple illumination directions.
\citet{weyrich2009fabricating} manufacture surface microgeometry on milled substrates to achieve target surface reflectance distributions; similarly, \citet{lu2025lumosx} exploit anisotropic reflection properties achievable through FDM printing parameters to encode directional information on object surfaces.

A complementary approach controls light by modifying the internal light delivery pathway rather than the shape.
\citet{willis2012printed} introduced ``Printed Optics,'' embedding custom light pipes and optical elements inside 3D-printed components to direct light to specific surface areas.
\citet{pereira2014computational} extend this concept to computational light routing, developing methods to route light through networks of 3D-printed optical fibers for sensing and display applications.

Several works have directly targeted lampshade design as a fabrication problem.
\citet{zhao2016printed} design 3D-printed perforated lampshades by distributing tubes of varying radii and orientations across an opaque shell, using 3D halftoning to project continuous grayscale images onto surrounding walls.
\citet{park2024easyshade} present a design tool for generating lampshades with stereographic-pattern projections via parametric modeling.

In contrast, our work targets translucent 3D-printed material, in which light undergoes refraction, scattering, and absorption throughout its volume.
We optimize the geometry of the internal cavity together with the number and positions of lights fed by optical fibers to achieve a uniformly luminous appearance of the lampshade.

\subsection{Design with Differentiable Rendering}
Differentiable rendering computes analytic derivatives of rendered images with respect to scene parameters, enabling gradient-based inverse design of geometry, materials, and lighting.

Modern differentiable rendering frameworks, notably Mitsuba~3 \citep{nimier2019mitsuba,jakob2022dr}, provide the infrastructure for such gradient computation.
\citet{li2018differentiable} established a foundational framework for differentiating through geometric boundaries via edge sampling, and adjoint methods~\citep{nimier2020radiative} enable scalable gradient computation over full light transport paths.
\citet{vicini2021path} introduced path replay backpropagation, which differentiates light paths using constant memory and linear time, and reparameterizations~\citep{loubet2019reparameterizing} handle discontinuous integrands arising from geometric visibility.

Differentiable rendering has been applied to a range of physical design problems.
\citet{tojo2023stealth} optimize the geometry of 3D objects to minimize radar cross-section while preserving their visible appearance.
\citet{gkioulekas2013inverse} use inverse rendering to estimate material properties within participating media.
\citet{nicolet2024inverse} leverage physically-based inverse rendering to optimize light projection profiles for tomographic volumetric additive manufacturing.
\citet{luci2024differentiable} develop a differentiable model of material spreading in inkjet printing to predict and optimize printed appearance.
\citet{lipp2024view} use differentiable rendering to optimize the placement of external light sources in architectural scenes to match a target illumination on surrounding surfaces.

Gradient-based optimization of geometry via differentiable rendering remains challenging, as rendering gradients are typically noisy and localized.
\citet{nicolet2021large} introduced Large Steps, a preconditioning scheme based on the mesh Laplacian that regularizes vertex updates and substantially improves the robustness of large-scale shape deformation.

Our work builds upon these foundations, applying Large Steps preconditioning to the internal cavity mesh under a fixed boundary, and leveraging Mitsuba~3 with physically calibrated participating media to achieve uniform luminance on a translucent lampshade with a complex free-form shape.

\section{Design Optimization}

This section presents our design optimization method (\autoref{fig:overview}). We assume that the material parameters of the translucent shade material have already been calibrated. This calibration is independent of the optimization and is described in \autoref{sec:calibration}.


\subsection{Problem Formulation}
We represent the lampshade as a composition of three components: an external mesh $M_\text{ext}$ defining the outer silhouette, an internal cavity mesh whose vertex positions form the design variable $\mathbf{x}_\text{mesh}$, and a set of lights whose positions form $\mathbf{x}_\text{light}$.
Given an external mesh, a target luminance image $I_\text{target}$, and material, camera, and light parameters, we optimize the design variables $\mathbf{x} = (\mathbf{x}_\text{mesh},\, \mathbf{x}_\text{light})$ to minimize an image-based objective rendered by Mitsuba~3:
\begin{equation}
\Phi(\mathbf{x}) = F(R(\mathbf{x})) + C(\mathbf{x}),
\end{equation}
where $R(\mathbf{x})$ denotes the rendered image under parameters $\mathbf{x}$, $F$ is the image fidelity term, and $C(\mathbf{x})$ aggregates collision and spacing penalties described in \autoref{sec:constraints}.
In this work, $F$ is defined as the mean squared error (MSE) between the rendered image and the target luminance image:
\begin{equation}
F(R(\mathbf{x})) = \frac{1}{|P|} \sum_{p \in P} \left( R_p(\mathbf{x}) - I_{\text{target},p} \right)^2,
\end{equation}
where $P$ is the set of all pixels in the image and $I_{\text{target}}$ is the target luminance image.
While we use MSE throughout this work, the fidelity term $F$ admits any differentiable loss, and alternative choices may better emphasize different aspects of the target luminance distribution.

\subsection{Scene Initialization}
\label{sec:initialization}

The initialization stage constructs the initial cavity mesh and determines the number and placement of lights.
A good initialization is important, as a poor starting point can cause the optimizer to become trapped in local minima or produce geometrically invalid configurations.

\paragraph{Cavity mesh.}
We initialize the internal cavity mesh by offsetting the external mesh inward by a user-specified margin $\delta$ (set to $5~\mathrm{mm}$ in our experiments).
Concretely, we evaluate the signed distance function of the external mesh on a regular volumetric grid, add $\delta$ to each sample to obtain the offset field, and extract its zero isosurface via Marching Cubes~\citep{lorensen1998marching}.
The resulting mesh is post-processed to remove isolated components and remeshed to ensure geometric regularity before optimization begins.

\begin{figure}[t]
  \centering
  \includegraphics[width=\linewidth]{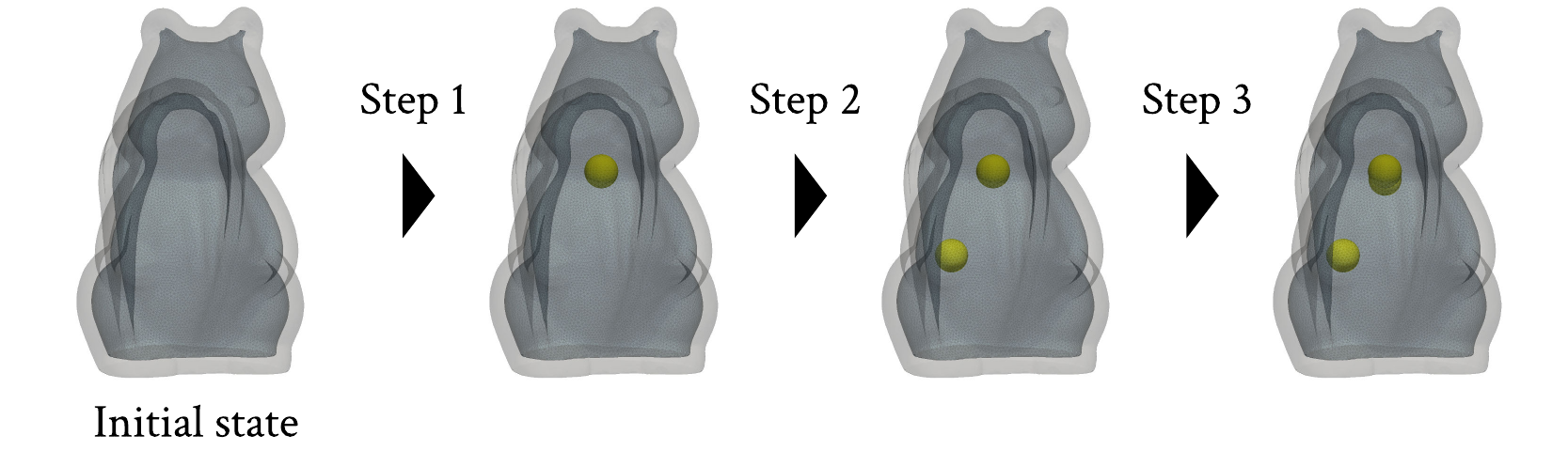}
  \caption{GreedyLightPlacement initialization. Starting from an empty configuration (left), lights are inserted one at a time by selecting the candidate position that most reduces $\Phi$, yielding an initial configuration of three lights (right).}
  \label{fig:greedy}
  \Description{Four panels illustrating the GreedyLightPlacement initialization for the squirrel lampshade. Leftmost panel labeled Initial state: the internal cavity with no lights placed. After Step 1: one yellow sphere light inserted in the upper body region. After Step 2: a second light added in the lower body. After Step 3: the final configuration of three lights distributed throughout the cavity.}
\end{figure}

\paragraph{Light placement.}
We determine the initial number and placement of lights via a GreedyLightPlacement initialization (\autoref{fig:greedy}).
We first place candidate positions on a regular three-dimensional grid spanning the bounding box of the internal mesh, with a grid spacing larger than the light diameter $2r$.
We then retain only those candidates that are strictly contained within the internal cavity, discarding any that fall outside or intersect its boundary.
Starting from an empty light set, we iteratively evaluate all remaining candidates and insert the one that most reduces $\Phi(\mathbf{x})$ given the lights already placed.
Insertion terminates when either no candidate yields a further reduction in $\Phi$ or the number of selected lights reaches a user-specified maximum $N_\text{max}$, at which point the light count $N$ and initial positions are fixed.
In our experiments, we set $N_\text{max} = 10$. Termination was always triggered by the first condition before the cap was reached.

\subsection{Optimization}

Mitsuba~3 computes gradients of $\Phi$ with respect to both $\mathbf{x}_\text{mesh}$ and $\mathbf{x}_\text{light}$ via differentiable rendering.
Light positions $\mathbf{x}_\text{light}$ are updated with the Adam optimizer~\citep{kingma2014adam}.
Mesh vertex positions $\mathbf{x}_\text{mesh}$ require additional care, as described below.

\paragraph{Mesh vertex updates.}
Applying Adam directly to mesh vertex positions is prone to producing local folds and self-intersections, because the stochastic rendering gradient is spatially localized and can drive individual vertices into geometrically invalid configurations.
To prevent this, we adopt the \emph{Large Steps} preconditioning scheme of \citet{nicolet2021large}, which regularizes vertex updates in the spectral domain of the mesh Laplacian:
\begin{equation}
\mathbf{x}_\text{mesh} \leftarrow \mathbf{x}_\text{mesh} - \eta\,(\mathbf{I} + \lambda \mathbf{L})^{-2}\,\frac{\partial \Phi}{\partial \mathbf{x}_\text{mesh}},
\end{equation}
where $\eta$ is the learning rate, $\mathbf{I}$ is the identity matrix, $\mathbf{L}$ is the cotangent Laplacian matrix of the mesh, and $\lambda > 0$ controls the strength of the smoothing (set to $\lambda = 25$ in our experiments).
The preconditioner $(\mathbf{I} + \lambda \mathbf{L})^{-2}$ attenuates high-frequency components of the gradient relative to low-frequency ones, so that the update preferentially deforms the mesh in smooth, large-scale modes.
A larger $\lambda$ increases this attenuation, resulting in smoother deformations at the cost of slower convergence toward fine geometric detail.

\subsection{Constraints}
\label{sec:constraints}
The constraint term $C(\mathbf{x})$ penalizes (i) intersections between the external and internal meshes, (ii) lights intersecting the internal mesh, and (iii) lights that are too close to each other:
\begin{equation}
C(\mathbf{x}) = \lambda_{\text{mm}} C_{\text{mm}}(\mathbf{x}) + \lambda_{\text{ml}} C_{\text{ml}}(\mathbf{x}) + \lambda_{\text{ll}} C_{\text{ll}}(\mathbf{x}),
\end{equation}
where $\lambda_{\text{mm}}$, $\lambda_{\text{ml}}$, and $\lambda_{\text{ll}}$ are the weights of the respective terms, set to $0.01$ in our experiments.
Each term is formulated as a squared hinge loss on a signed distance function (SDF) or Euclidean distance.
The mesh-mesh term $C_{\text{mm}}$ penalizes vertices of the internal mesh that approach or cross the external mesh:
\begin{equation}
C_{\text{mm}}(\mathbf{x}) = \sum_{\mathbf{v}_i \in \mathbf{x}_\text{mesh}} \max\left(\mathrm{SDF}(\mathbf{v}_i, M_{\text{ext}}) + d_{\text{mm}}, 0\right)^2,
\end{equation}
where $\mathbf{v}_i \in \mathbb{R}^3$ is a vertex position of the internal mesh $\mathbf{x}_\text{mesh}$, $\mathrm{SDF}(\cdot, M_{\text{ext}})$ is the signed distance function of the external mesh $M_{\text{ext}}$, and $d_{\text{mm}}$ is an offset margin.
The mesh-light term $C_{\text{ml}}$ penalizes lights that approach or cross the internal mesh:
\begin{equation}
C_{\text{ml}}(\mathbf{x}) = \sum_{i=1}^{N} \max\left(\mathrm{SDF}(\mathbf{l}_i, \mathbf{x}_\text{mesh}) + r + d_{\text{ml}}, 0\right)^2,
\end{equation}
where $\mathbf{l}_i \in \mathbb{R}^3$ is the position of the $i$-th light, $N$ is the total number of lights, $\mathrm{SDF}(\cdot, \mathbf{x}_\text{mesh})$ is the signed distance function of the internal mesh $\mathbf{x}_\text{mesh}$, $r$ is the light radius (set to $r = 6~\mathrm{mm}$ in our experiments), and $d_{\text{ml}}$ is an offset margin.
The light-light term $C_{\text{ll}}$ penalizes pairs of lights that are too close to or intersect one another:
\begin{equation}
C_{\text{ll}}(\mathbf{x}) = \sum_{1 \le i < j \le N} \max\left(2r + d_{\text{ll}} - \lVert \mathbf{l}_i - \mathbf{l}_j \rVert_2, 0\right)^2,
\end{equation}
where $d_{\text{ll}}$ is an offset margin.
In our experiments, $d_{\text{mm}} = d_{\text{ml}} = d_{\text{ll}} = 1~\mathrm{mm}$.
\section{Fabrication}

The lampshade is fabricated by fused deposition modeling (FDM) using translucent polycarbonate (PC) filament.
We use pseudo sphere lights (described below) suspended at their optimized positions within the cavity via printed struts.
To accommodate the pseudo sphere lights inside the closed cavity, the shade is split horizontally and printed as two separate halves that are joined during assembly.

\begin{figure}[t]
  \centering
  \includegraphics[width=\linewidth]{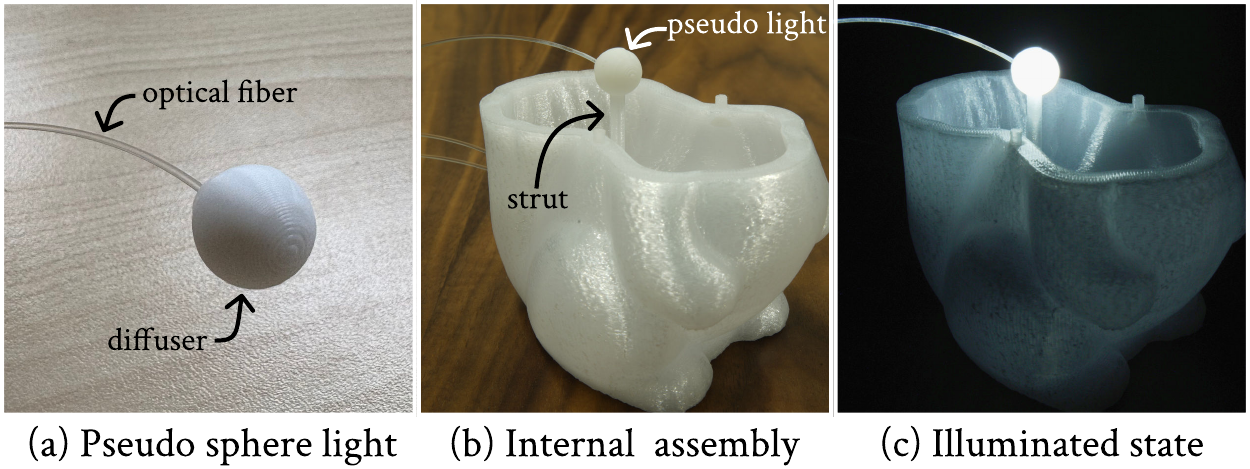}
  \caption{Pseudo sphere light fabrication. (a) Pseudo sphere light: a PLA spherical diffuser ball coupled to a thin optical fiber. (b) Internal assembly: the 3D-printed lampshade half with a pseudo sphere light mounted at its optimized position via a printed strut. (c) A lampshade half in the illuminated state.}
  \Description{Three panels showing the fabrication process: (a) a close-up photograph of the pseudo sphere light consisting of a small spherical diffuser ball attached to a thin optical fiber; (b) a photograph of the 3D-printed lampshade half with the pseudo sphere light mounted on a printed strut at the optimized position; (c) a photograph of the lampshade half in a dark environment, showing the pseudo sphere light glowing on its strut inside the open cavity, with the translucent walls illuminated by the transmitted light.}
  \label{fig:fabrication}
\end{figure}

\paragraph{Pseudo sphere light.}
Conventional light sources such as LED modules are too bulky to be positioned with precision within a compact internal cavity.
We instead employ a pseudo sphere light consisting of a spherical diffuser ball 3D-printed in PLA, coupled to a thin optical fiber (\autoref{fig:fabrication}(a)).
The slender profile of the optical fiber allows it to be routed through printed struts to reach positions that a conventional light source could not occupy.
The spherical diffuser produces a near-isotropic emission profile that approximates the sphere light model assumed in simulation, reducing the discrepancy between the simulated and fabricated appearance.

\paragraph{Struts.}
Since the pseudo sphere lights must be suspended at their optimized positions within the cavity, each is supported by a strut that routes the optical fiber to the diffuser (\autoref{fig:fabrication}(b)).
The struts are printed as part of the shade halves.

\section{Material Parameter Calibration}
\label{sec:calibration}

To close the gap between simulation and the fabricated result, we calibrate the material parameters governing the optical properties of the translucent shade material: the surface roughness and the volumetric scattering parameters, namely the extinction coefficient $\sigma_\mathrm{t}$, the albedo, and the anisotropy parameter $g$ of the Henyey-Greenstein phase function~\citep{henyey1941diffuse}.
An incorrect value for any of these parameters causes the simulated appearance to diverge from the fabricated result, undermining the validity of the optimization.
Since these parameters vary with the choice of material and printer, each user can run the calibration procedure for their own setup.

\subsection{Light Intensity}
Accurate material calibration requires knowing the absolute intensity of the light source used during the measurement procedure, which we therefore establish first.
To enable precise intensity modeling, we fit the light source with a black plate perforated with a square aperture, treating the assembly as a square area light.
We calibrate its intensity by photographing it directly and matching the result against a rendering of an identical virtual scene: we construct a virtual scene reproducing the camera and light placement used in the photograph, render it with a candidate intensity value, and optimize this value so that the mean pixel intensity within the aperture region matches that in the photograph.
Because the light source is bright, the photograph must be taken at a lower ISO and faster shutter speed than those used during the subsequent calibration, to avoid saturation.
The calibrated intensity is tied to these specific exposure settings. When the light source is subsequently used under the calibration exposure conditions, the intensity value must be rescaled by the corresponding exposure ratio.

\subsection{Surface Roughness Parameters}
\begin{figure}[t]
  \centering
  \includegraphics[width=\linewidth]{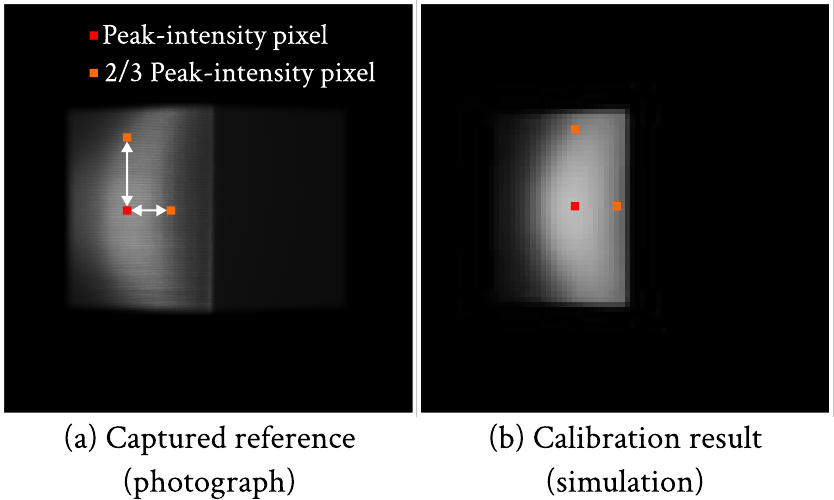}
  \caption{Roughness calibration. (a) Reference photograph of the specular reflection on the opaque cube; the peak-intensity pixel (red) and the pixels at $2/3$ peak intensity (orange) define the decay distances along the two surface tangent directions. (b) Simulation rendering after roughness calibration; the decay distances agree with those in~(a).}
  \Description{Two panels showing the roughness calibration procedure. (a) A photograph of the specular reflection on the opaque black cube, with a red dot marking the peak-intensity pixel and orange dots marking the pixels at $2/3$ peak intensity along each of the two surface tangent directions, defining the decay distances used as the calibration objective. (b) A simulation rendering of the same scene after roughness calibration, showing matching decay distances along both tangent directions.}
  \label{fig:roughness}
\end{figure}

We calibrate the surface roughness of the printed material using a cube printed in a fully opaque material.
Opacity is essential at this stage: in a translucent material, the observed reflection pattern would depend not only on surface roughness but also on the volumetric scattering parameters, which have not yet been calibrated, confounding the measurement.
In our experiments, the shade material is polycarbonate (PC), and we use a cube printed in black PLA for this calibration.
We illuminate one face of the cube obliquely with the light source and photograph the resulting specular reflection pattern, then construct a matching virtual scene (\autoref{fig:roughness}).
The material roughness is modeled with two parameters $\alpha_u$ and $\alpha_v$, one per surface tangent direction, which we optimize jointly.
The objective is defined in terms of the \emph{decay distance}: starting from the brightest pixel of the reflection pattern, the decay distance along a given direction is the number of pixels at which the intensity falls to a specified fraction of its peak value (set to $2/3$ in our experiments).
We compute the decay distances along both tangent directions in the photograph and the rendering, and minimize the sum of squared differences between the photographed and rendered decay distances, yielding the calibrated roughness.
The calibrated roughness parameters are $\alpha_u = 0.091$ and $\alpha_v = 0.434$.

\subsection{Volumetric Scattering Parameters}
\begin{figure*}[t]
  \centering
  \includegraphics[width=\textwidth]{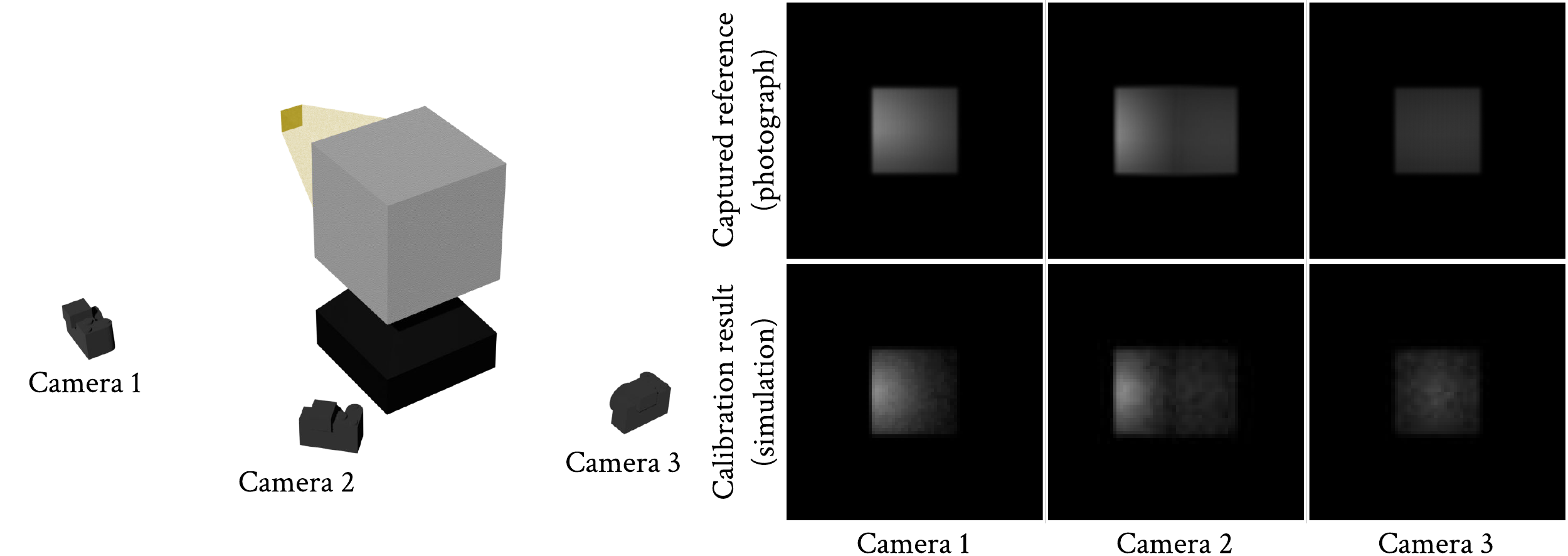}
  \caption{Calibration of the volumetric scattering parameters ($\sigma_\mathrm{t}$, albedo, and $g$). Left: measurement setup in which a translucent polycarbonate cube is illuminated by the light source and photographed from three viewpoints: from the side (Camera~1), from an oblique angle (Camera~2), and from directly behind the light source (Camera~3). Right: captured reference photographs (top row) and corresponding simulation renderings after joint calibration (bottom row) for each viewpoint.}
  \Description{Left side: a 3D diagram of the calibration setup showing a white cube illuminated by a small light source above, with three camera icons labeled Camera 1, Camera 2, and Camera 3 positioned at different viewing angles around the cube. Right side: a three-by-two grid showing captured photographs of the illuminated translucent cube (top row) and matching simulation renderings after parameter calibration (bottom row) from each of the three camera viewpoints.}
  \label{fig:other-params}
\end{figure*}

The remaining three parameters describe light transport within the translucent shade material: the extinction coefficient $\sigma_\mathrm{t}$, which governs the rate at which light is attenuated as it travels through the medium; the albedo, which governs the fraction of attenuated light that is scattered rather than absorbed; and the anisotropy parameter $g$ of the Henyey-Greenstein phase function, which governs the directional bias of scattering events.
To calibrate these three parameters jointly, we print a cube of the translucent shade material (PC in this work) and place the light source directly against one of its faces.
We photograph the cube from three viewing directions: from the side, from an oblique angle, and from directly behind the light source such that the light, the cube, and the camera are collinear (\autoref{fig:other-params}). We use these photographs as target images.
We then construct a matching virtual scene and jointly optimize $\mathbf{k} = (\sigma_\mathrm{t},\, \text{albedo},\, g)$ by minimizing the following weighted objective~\citep{gkioulekas2013inverse}:
\begin{equation}
\Phi(R(\mathbf{k})) = \frac{1}{|P_\text{mask}|} \sum_{p \in P_\text{mask}} \omega_p \left( R_p(\mathbf{k}) - I_{\text{target},p} \right)^2,
\end{equation}
where $P_\text{mask}$ is the set of pixels within a slightly eroded mask applied to each view to exclude boundary regions, and
\begin{equation}
\omega_p = \frac{1}{\max(c,\, I_{\text{target},p}^{\alpha})}
\end{equation}
with $c = 0.01$ and $\alpha = 3$.
The per-pixel weight $\omega_p$ up-weights darker pixels, giving the optimizer balanced sensitivity across the full luminance range rather than letting bright regions dominate the loss.
The calibrated volumetric scattering parameters are $\sigma_\mathrm{t} = 0.273~\mathrm{mm}^{-1}$, albedo $= 0.870$, and $g = 0.481$.

\section{Experiments}

\subsection{Setup}
\label{sec:setup}

We validate our method using a squirrel mesh as the lampshade geometry.
The target appearance is uniform luminance over the shade silhouette as seen from the camera, encoded as a constant target luminance image $I_\text{target}$ with value $0.4$ over the silhouette mask and $0.0$ outside.

All experiments are conducted on a single NVIDIA H100 GPU.
Optimization uses a rendering resolution of $150 \times 100$ pixels at 1024 samples per pixel (spp), a learning rate of $1.0$, and runs for 30{,}000 iterations, taking approximately 10~h per run.
The final design is taken as the iterate with the lowest objective value $\Phi$ over all iterations.
All material parameters are obtained through the calibration procedure described in \autoref{sec:calibration} prior to optimization.

\subsection{Design Optimization Results}
\label{sec:result}

\begin{figure}[t]
  \centering
  \includegraphics[width=\linewidth]{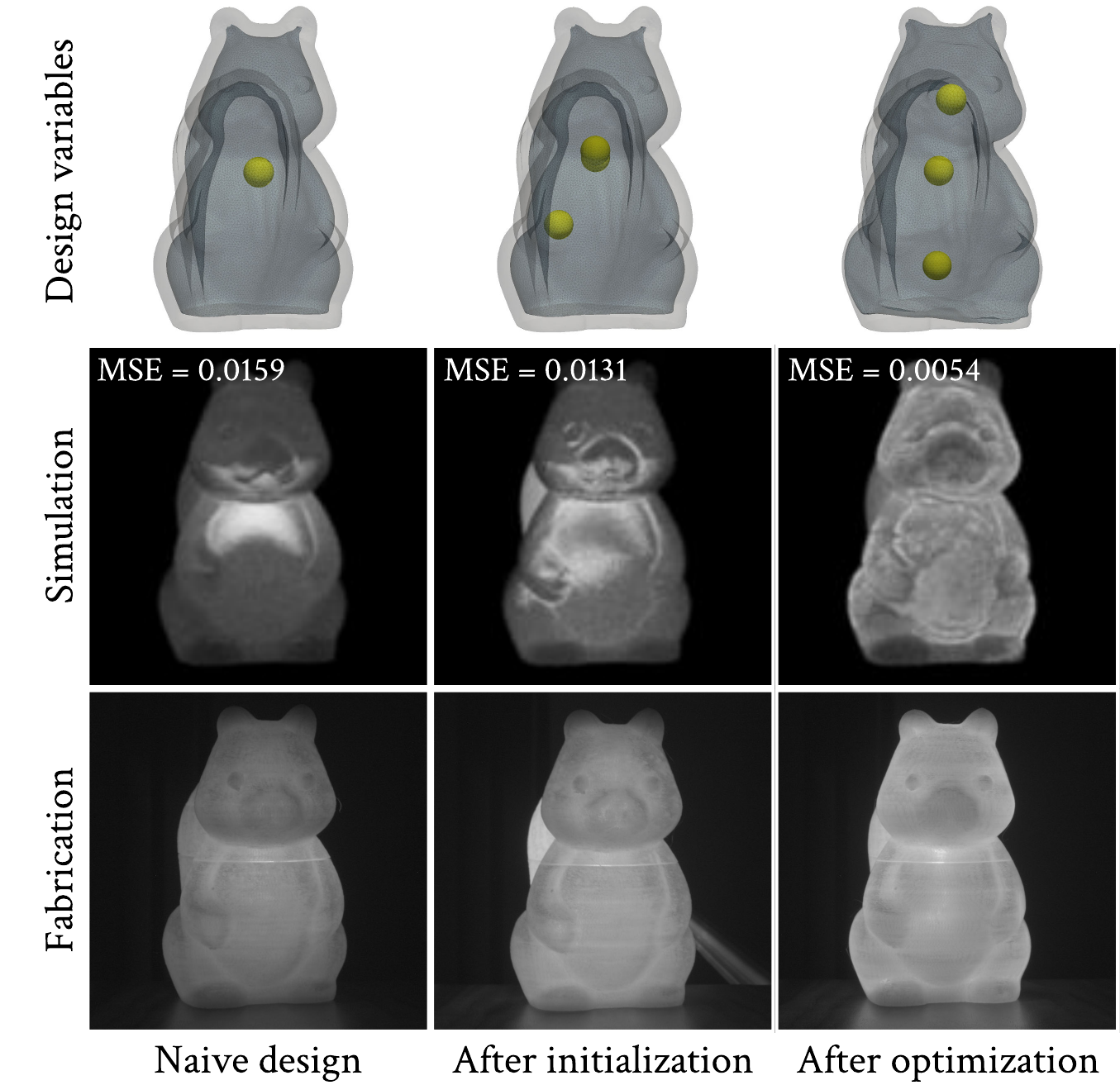}
  \caption{Progression from naive design to the fully optimized result for the squirrel lampshade (columns, left to right: naive single-light design, after light placement initialization, after full gradient-based optimization). Top row: internal scene showing the cavity mesh and light positions. Middle row: simulation renderings. Bottom row: photographs of the corresponding fabricated lampshades. MSE with respect to the target appearance image decreases from $0.0159$ (naive) to $0.0131$ (after initialization) to $0.0054$ (after optimization).}
  \Description{A three-by-three grid of images. Columns from left to right: naive single-light design, after light placement initialization, after full gradient-based optimization. Top row: internal scene showing the cavity mesh and light positions. Middle row: simulation renderings of the squirrel lampshade, showing luminance concentrated near the neck in the naive and initialized designs and substantially more uniform after optimization. Bottom row: photographs of the corresponding fabricated lampshades showing the same progression.}
  \label{fig:result}
\end{figure}

\autoref{fig:result} shows the rendered appearance and the internal scene geometry at each stage of the pipeline.
As a naive baseline, we place a single light at the center of the shape and use the initial cavity mesh without optimization.
The greedy initialization places three lights within the internal cavity, reducing the MSE from $0.0159$ (naive single-light design) to $0.0131$.
At this stage, two lights are clustered near the neck region and one is placed toward the body, resulting in high luminance near the neck and insufficient luminance in the head.
Gradient-based optimization then jointly refines the light positions and cavity geometry, further reducing the MSE to $0.0054$, a reduction substantially larger than that achieved by initialization alone.
The fabricated lampshade achieves a visually uniform appearance, consistent with the optimization target.

\subsection{Ablation Studies}
\label{sec:ablation}

\subsubsection{Effect of Material Parameter Calibration}
\label{sec:ablation-calib}

\begin{figure}[t]
  \centering
  \includegraphics[width=\linewidth]{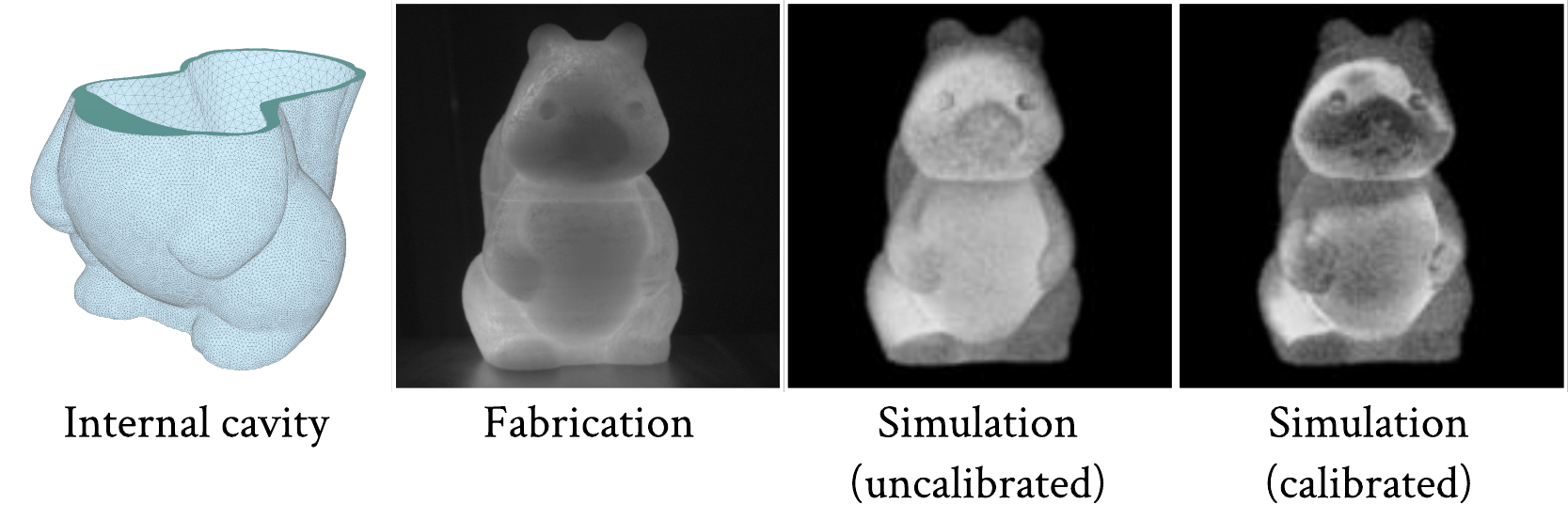}
  \caption{Effect of material parameter calibration. From left to right: internal cavity geometry, photograph of the fabricated lampshade, simulation rendered with uncalibrated parameters, and simulation rendered with calibrated parameters. The calibrated simulation more closely reproduces the fabricated appearance than the uncalibrated simulation.}
  \Description{Four images shown side by side for a single squirrel lampshade design. From left to right: the internal cavity mesh, a photograph of the fabricated lampshade, a simulation rendering with uncalibrated material parameters, and a simulation rendering with calibrated material parameters. The calibrated rendering more closely matches the fabricated photograph than the uncalibrated rendering.}
  \label{fig:ablation-calib}
\end{figure}

To assess the importance of material parameter calibration, we fabricate a squirrel lampshade and compare its appearance against renderings produced with calibrated and uncalibrated parameters (\autoref{fig:ablation-calib}).
The uncalibrated simulation fails to reproduce the fabricated appearance, whereas the calibrated simulation more closely matches it, confirming that accurate material parameters are necessary for the simulation to correctly predict the physical result.

\subsubsection{Effect of Geometric Constraints}
\label{sec:ablation-constraints}

\begin{figure}[t]
  \centering
  \includegraphics[width=\linewidth]{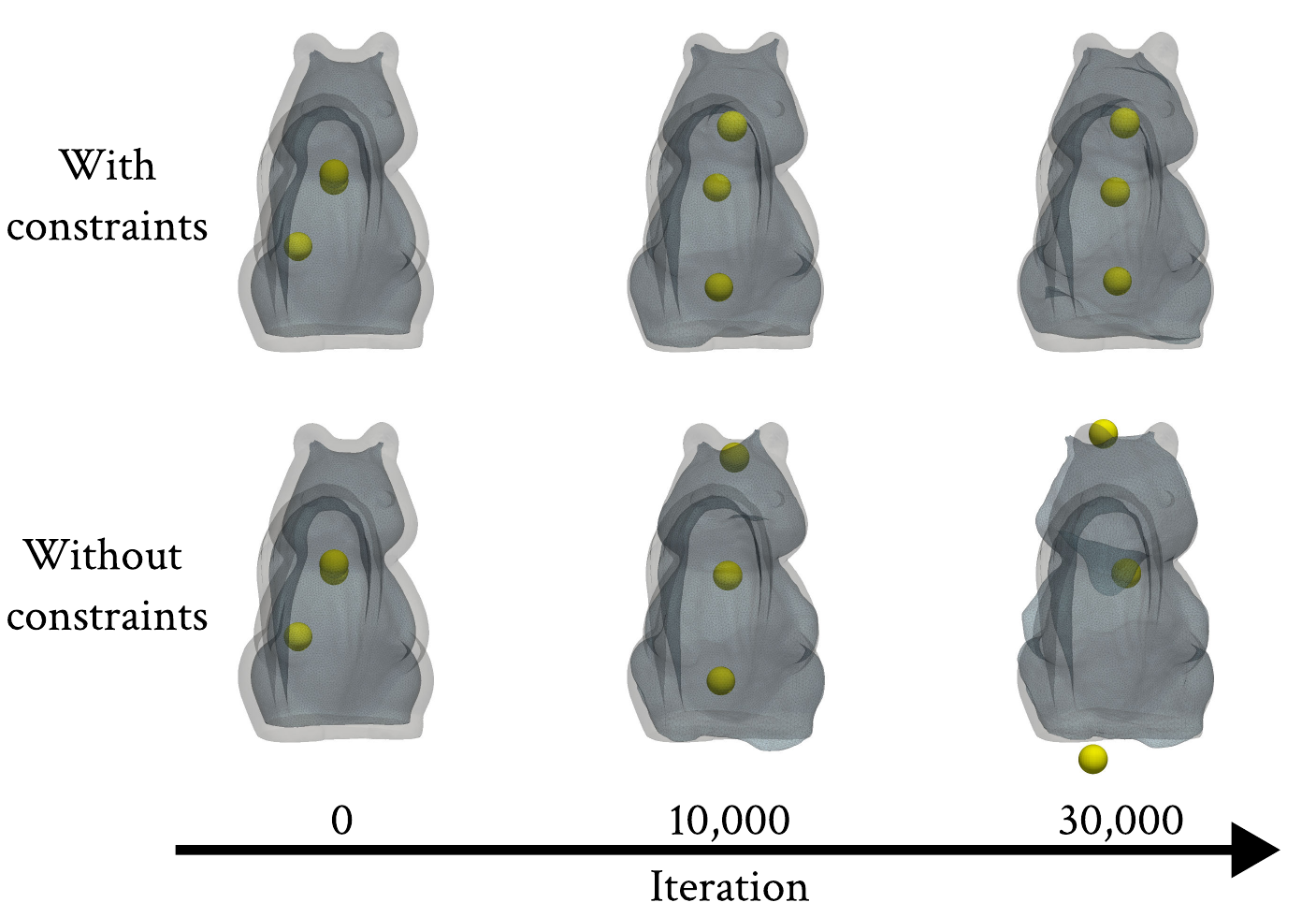}
  \caption{Effect of geometric constraints on optimization (columns: iterations 0, 10{,}000, and 30{,}000). Top row: with the constraint term $C(\mathbf{x})$ enabled, the internal mesh and lights remain within the valid cavity throughout. Bottom row: without $C(\mathbf{x})$, by iteration 10{,}000 the internal mesh slightly penetrates the external mesh and a light becomes partially embedded in the internal mesh. By iteration 30{,}000 the violations worsen and a light has exited entirely through the external mesh.}
  \Description{A two-by-three grid of 3D scene visualizations showing the squirrel lampshade interior at iterations 0, 10000, and 30000. Top row: with the geometric constraint term enabled, the internal mesh and lights remain within the valid cavity throughout optimization. Bottom row: without the constraint term, by iteration 10000 the internal mesh slightly crosses the external mesh boundary and a light is partially embedded in the internal mesh. By iteration 30000 the violations are more severe and a light has passed entirely through the external mesh.}
  \label{fig:ablation-constraints}
\end{figure}

The constraint term $C(\mathbf{x})$ is designed to prevent three types of geometric violation: the internal mesh penetrating the external mesh ($C_\text{mm}$), lights becoming embedded in the mesh ($C_\text{ml}$), and lights clustering too close together ($C_\text{ll}$).
To assess the necessity of these penalties, we run optimization with $C(\mathbf{x})$ removed entirely.
As shown in \autoref{fig:ablation-constraints}, the unconstrained optimization progressively produces geometrically invalid configurations: by iteration 10{,}000, the internal mesh slightly penetrates the external mesh and a light becomes partially embedded in the internal mesh. By iteration 30{,}000, the violations worsen and a light has exited entirely through the external mesh.
These violations make the design physically unrealizable: an internal mesh that penetrates the external surface or a light that protrudes outside the cavity cannot be fabricated.
The constraint term is therefore not merely a regularizer but a prerequisite for the optimization to produce a valid, physically realizable design.

\section{Demonstrations}

\begin{figure*}[p]
  \centering
  \includegraphics[width=\textwidth, height=0.9\textheight, keepaspectratio]{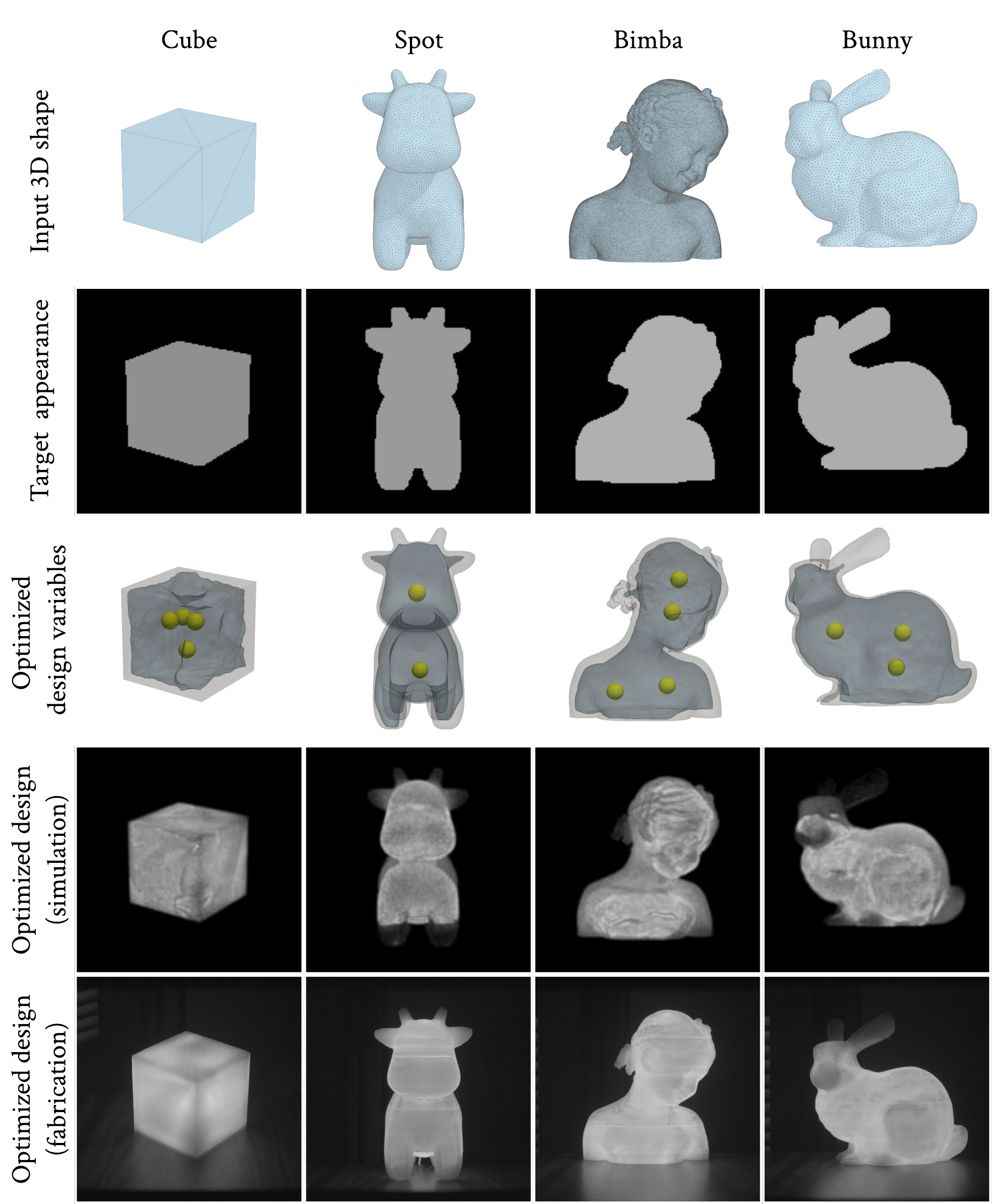}
  \caption{Demonstrations on four shapes (columns, left to right: cube, Spot, Bimba, and Stanford Bunny). Rows from top to bottom: input 3D mesh; target appearance image; optimized internal scene showing the deformed cavity mesh and light positions; simulation rendering of the optimized design; and photograph of the fabricated lampshade. All shapes use a uniform luminance target ($I_\mathrm{target}$ constant within the silhouette; value varies per shape).}
  \Description{A five-by-four grid of images. The four columns correspond to four lampshade shapes: cube, Spot, Bimba, and Stanford Bunny. The five rows show from top to bottom: the input 3D mesh, the target appearance image, the optimized internal scene with lights shown as yellow spheres inside the translucent cavity, the simulation rendering of the optimized design, and a photograph of the fabricated lampshade.}
  \label{fig:demonstrations}
\end{figure*}

To demonstrate the generality of our method, we apply it to four shapes of varying geometric complexity, each with uniform luminance as the optimization target.
All material parameters are obtained through the calibration procedure described in \autoref{sec:calibration}.
Each optimization run uses the same settings as described in \autoref{sec:setup} and takes approximately 10~h.

\paragraph{Cube.}
The cube provides a simple baseline with flat faces and sharp edges.
The optimization places four lights within the internal cavity, producing uniform luminance across the flat faces.
Corners and edges exhibit slightly lower luminance than the face centers, though the overall appearance remains close to uniform.
The target pixel value is $I_\text{target} = 0.4$ within the silhouette.

\paragraph{Spot.}
Spot is a compact organic character with a rounded body and thin extremities including limbs, horns, and ears.
The optimization places two lights: one in the head and one in the body.
The main body achieves relatively uniform luminance. The thin extremities remain dark, as the interior volume in these regions is too narrow to accommodate a sphere light.
The target pixel value is $I_\text{target} = 0.5$ within the silhouette.

\paragraph{Bimba.}
Bimba is a classical humanoid bust with detailed surface features including facial contours, eye sockets, and flowing hair.
The optimization places four lights: two in the head region and two in the body.
Despite the geometric complexity, the fabricated result achieves relatively uniform luminance across the shade.
The target pixel value is $I_\text{target} = 0.8$ within the silhouette.

\paragraph{Stanford Bunny.}
The Stanford Bunny is a widely used benchmark shape with complex surface curvature.
The optimization places three lights within the internal cavity, producing relatively uniform luminance over the main body.
As with Spot, the ears remain insufficiently illuminated due to their narrow geometry.
The target pixel value is $I_\text{target} = 0.75$ within the silhouette.

\autoref{fig:demonstrations} shows the results for all four demonstrations.

\section{Discussion}

\subsection{Limitations}

\paragraph{Target appearance specification.}
The target appearance image $I_\mathrm{target}$ is currently specified manually by the user.
The absolute pixel values directly govern the luminance level the optimizer attempts to achieve, and a poorly chosen value can bias the solution toward a luminance level that is physically unrealizable given the available light power or, conversely, that leaves significant luminous flux unused.
Potential remedies include designing a scale-invariant objective that is insensitive to the absolute luminance level, thereby focusing the optimizer on the spatial distribution rather than the absolute value, or jointly optimizing the light intensity alongside the geometric design variables so that a feasible luminance level is determined automatically.

\paragraph{Simulation-to-fabrication gap.}
Although we calibrate the key material parameters prior to optimization (\autoref{sec:calibration}), several sources of discrepancy between simulation and the physical fabrication remain.
First, the calibrated material parameters may not perfectly reflect the actual optical properties of the printed shade.
In particular, surface roughness is estimated using a cube printed in black PLA rather than the translucent shade material (PC), since the opacity of the calibration target is required to isolate surface reflection. As a result, the estimated roughness may not accurately represent the PC surface finish.
Developing a method to estimate surface roughness directly from translucent materials would be a valuable direction.
Second, we model the lights as an ideal sphere emitter with uniform surface radiance. In practice, the emission profile of the spherical diffuser ball deviates from this ideal, introducing a discrepancy between the simulated and fabricated appearance.
Third, fused deposition modeling (FDM) introduces anisotropic optical properties arising from the layer-by-layer deposition process. While our model accounts for some anisotropy, the full directional variation in transmittance and scattering produced by the layered structure is more complex than this parameterization can capture.
The Henyey-Greenstein phase function may lack the expressiveness to fully capture the directional scattering of FDM-printed materials, and a more accurate model could yield different optimal light placements.
Finally, the printed struts that hold the lights at their optimized positions cast shadows and partially occlude light paths within the cavity. These are not included in the differentiable rendering stage and can produce localized dark regions in the fabricated result.

\paragraph{Thin geometric features.}
A recurring limitation across shapes is that thin geometric features such as ears, horns, and limbs consistently receive insufficient illumination.
This arises because the inward-offset internal mesh further reduces the already narrow interior volume in thin regions, and the greedy initialization requires each candidate sphere light of radius $r = 6~\mathrm{mm}$ to be strictly contained within the internal mesh.
In sufficiently thin regions, no grid candidate satisfies this containment constraint, so no light is placed there and the deficit persists through optimization.
Addressing this would require either an adaptive light initialization strategy or a smaller-diameter diffuser. The current radius is a fabrication choice, not a fundamental limit, and could be reduced toward the fiber's thickness to reach thinner regions.

\paragraph{Expressiveness of design variables.}
Our method optimizes the internal cavity geometry and light positions, but this parameterization has inherent limits.
For geometries with very thin or deeply concave regions, neither repositioning the lights nor deforming the internal cavity may provide sufficient control to redirect adequate luminous flux into every part of the shade.
Incorporating additional design degrees of freedom, such as spatially varying material properties or auxiliary internal reflectors, would be necessary to handle such cases.

\subsection{Future Work}

\paragraph{Non-uniform luminance targets.}
This work focuses on uniform luminance, but the method accepts an arbitrary target luminance image, and extending to non-uniform targets requires no fundamental change to the formulation.
However, achieving reliable reproduction of spatially structured patterns places higher demands on simulation-to-fabrication fidelity, as even small simulation errors can distort the intended spatial pattern.
Improving the accuracy of the simulation model would therefore be a prerequisite for this extension.
We leave experimental validation of non-uniform targets to future work.

\paragraph{Multi-view optimization.}
The current formulation optimizes a single front-facing rendered image as the objective.
In practice, a lampshade is observed from multiple directions simultaneously, and the target luminance distribution as seen from one viewpoint does not guarantee the same distribution from others.
Extending the objective to a weighted combination of renderings from multiple viewpoints would enable the optimizer to account for the full angular appearance of the shade.

\paragraph{Color appearance.}
All experiments in this work use grayscale luminance as the optimization target.
Extending the method to full-color rendering would open the possibility of designing lampshades that emit spatially varying chromaticity, for example by producing color gradient effects or matching a prescribed color pattern on the shade surface.
Differentiable path tracers such as Mitsuba~3 natively support multi-spectral rendering, so the primary requirements are a multi-channel target image and an appropriate colorimetric loss function.

\paragraph{Optimization evaluation.}
While we quantitatively assess the optimization via simulation MSE, the fabricated results are evaluated qualitatively.
Photometric measurements of the fabricated appearance would enable a more rigorous evaluation of how well the optimized design transfers to the physical object.
Additionally, evaluating how the illumination changes under perturbations of the optimized light positions and internal cavity geometry would characterize the robustness of the optimized configurations and clarify the contribution of gradient-based refinement.

\paragraph{Material distribution optimization.}
A promising direction for increasing design expressiveness is to include the spatial distribution of material properties as additional design variables.
For instance, locally embedding high-reflectance or high-scattering material in selected regions of the internal surface, such as a diffusely reflective coating applied to areas that are geometrically difficult to illuminate from the available light positions, would allow light to be redirected into regions that neither geometry deformation nor light repositioning can adequately reach.
Jointly optimizing such a spatial material map with the internal cavity geometry and light placement, using differentiable rendering to propagate gradients through the volumetric scattering model, would substantially broaden the range of achievable luminance distributions.

\section{Conclusion}

We have presented a computational design method for 3D-printable lampshades with complex free-form shapes.
Given an arbitrary shape, our method determines the internal cavity geometry and the number and placement of lights in two stages: an initialization stage that establishes the initial design variables, followed by gradient-based optimization via differentiable rendering that jointly refines the cavity geometry and light positions subject to geometric constraints.
A material calibration procedure improves the agreement between simulation and the fabricated appearance.
We validate the approach by fabricating five lampshades of varying geometric complexity and demonstrate that the fabricated results achieve uniform luminance consistent with the optimization target.
We hope the insights and techniques developed in this work will be useful to future research on fabrication design via differentiable rendering.

\begin{acks}
This work was supported in part by JSPS KAKENHI Grant Number JP26K14879.
This research was conducted using the Supermicro ARS-111GL-DNHR-LCC and FUJITSU Server PRIMERGY CX2550 M7 (Miyabi) at Joint Center for Advanced High Performance Computing (JCAHPC).
The Spot model is courtesy of Keenan Crane~\citep{crane2013robust}, the Bimba model is courtesy of the AIM@SHAPE Shape Repository, and the Stanford Bunny model is courtesy of the Stanford University Computer Graphics Laboratory.
\end{acks}
\bibliographystyle{ACM-Reference-Format}
\bibliography{meta/references}

\end{document}